%% file: paper.tex
\documentclass{sigcomm-alternate}
\usepackage{graphicx,amsmath,amssymb,url,subfigure}
\begin{document}

\title{The Internet AS-Level Topology: \\ Three Data Sources and One Definitive Metric}

\numberofauthors{1}

\author{
    Priya Mahadevan \\ UCSD \and
    Dmitri Krioukov \\ CAIDA \and
    Marina Fomenkov \\ CAIDA \and
    Bradley Huffaker \\ CAIDA \and
    Xenofontas Dimitropoulos \\ Georgia Tech \and
    kc claffy \\ CAIDA \and
    Amin Vahdat \\ UCSD \and
    {\normalsize \{pmahadevan,vahdat\}@cs.ucsd.edu, \{dima,marina,brad,fontas,kc\}@caida.org}
}

\maketitle

\begin{abstract}
\noindent
We calculate an extensive set of characteristics for Internet AS
topologies extracted from the three data sources most frequently used
by the research community: traceroutes, BGP, and WHOIS.  We discover that
traceroute and BGP topologies are similar to one another but differ
substantially from the WHOIS topology. Among the widely considered
metrics, we find that the {\em joint degree distribution} appears to
fundamentally characterize Internet AS topologies as well as narrowly
define values for other important metrics. We discuss the interplay
between the specifics of the three
data collection mechanisms and the resulting topology views. In particular, we show
how the data collection peculiarities explain differences in the
resulting joint degree distributions of the respective topologies.
Finally, we release to the community the input topology
datasets, along with the scripts and output of our calculations. This
supplement should enable researchers to validate their models against
real data and to make more informed selection of topology data sources
for their specific needs.
\end{abstract}




\section{Introduction}
\label{sec:intro}
\input{intro}

\section{Data Sources}
\label{sec:data}
\input{data}

\section{Topology Characteristics}
\label{sec:characteristics}
\input{characteristics}

\section{Conclusion}
\label{sec:conclusion}
\input{conclusion}

\section{Acknowledgments}
\input{ack}

\input{paper.bbl}
\end{document}

%% file: intro.tex
Internet topology analysis and modeling has attracted substantial attention
recently~\cite{FaFaFa99,WilRevisited,
TaGoJaShWi02,LiAlWiDo04,BuTo02,JaRoTo04,ZhoMo04}\footnote{We
intentionally avoid citing the statistical physics literature, where the number of
publications dedicated to the subject has exploded. For an introduction and
references see~\cite{DorMen-book03}.} because the Internet's topological properties and their evolution are
cornerstones of many practical and theoretical network research agendas.
Routing, performance of applications and protocols, robustness of the network under
attack, {\it etc.}, all depend on network topology. Since obtaining realistic topology
data is crucial for the above agendas, researchers have
focused on a variety of measurement techniques to capture the
Internet's topology.

Various sources of Internet topology data obtained using different
methodologies yield substantially different
topological views of the Internet. Unfortunately, many researchers either rely
only on one data source, sometimes outdated or incomplete, or mix disparate
data sources into one topology. To date, there has been little attempt to provide a
detailed analytical comparison of the most important properties of topologies
extracted from the different data sources.

Our study fills this gap by analyzing and explaining topological properties
of Internet AS-level graphs extracted from the three commonly-used data sources:
(1)~traceroute measurements~\cite{skitter}; (2)~BGP~\cite{routeviews}; and
(3)~the WHOIS database~\cite{irr}.

This work makes three key contributions to the field of topology research:
\begin{enumerate}
\item We calculate a range of topology metrics considered in the
networking literature for the three sources of data. We reveal the peculiarities of each data source and the
resulting interplay between artifacts of data collection and the key properties
of the {\em joint degree distributions} of the derived graphs.

\item We analyze the interdependencies among an array of topological features
and observe that the {\em joint degree distributions} of the graphs define
other crucial topological characteristics.

\item To promote and simplify further analysis and discussion, we
release~\cite{comp-anal} the following data and results to the community:
a)~the AS-graphs representing the topologies extracted from the raw data sources;
b)~the full set of data plots (many not included in the paper) calculated for
all graphs;
c)~the data files associated with the plots, useful for researchers
looking for other summary statistics or for direct comparisons with
empirical data;
and d)~the scripts and programs we developed for our calculations.
\end{enumerate}

We organize this paper as follows. Section~\ref{sec:data} describes our data
sources and how we constructed AS-level graphs from these data.
In Section~\ref{sec:characteristics} we present the set of topological
characteristics calculated from our graphs and explain what they measure and
why they are important. We conclude in Section~\ref{sec:conclusion} with a
summary of our findings.

%% file: data.tex
\subsection{Constructing AS graphs}
We used the following data sources to construct AS-level graphs of the Internet: traceroute
measurements, BGP data, and the WHOIS database.
We make all of our
constructed graphs publicly available~\cite{comp-anal}.

{\bf BGP} (Border Gateway Protocol)~\cite{bgp} is the protocol for
routing among ASes in the Internet.
RouteViews~\cite{routeviews} collects BGP routing tables using~7 collectors, 5 of which
are located in the USA, 1 in the UK and 1 in Japan.
Each collector has a number of globally placed peers (or vantage
points) that collect BGP messages from which we can infer the AS topology.
RouteViews archives both static snapshots of the BGP routing tables and dynamic
BGP data in the form of BGP message dumps
(updates and withdrawals). Therefore, we derive two types of
graphs from the BGP data for the same month of March~2004: one from the static
tables ({\bf BGP tables}) and one from the updates ({\bf BGP updates}).
We create the BGP tables graph using data from the collector
{\it route-views.oregon-ix.net} as it gathers data from the largest number of peers---68.
For the BGP updates graph, we choose the collector {\it route-views2.oregon-ix.net},
which uses 40 peers to collect data, since at the time of this research
{\it route-views.oregon-ix.net} did not collect BGP updates. The data contains AS-sets~\cite{bgp},
that is, lists of ASes with unknown interconnection structures.
For both BGP tables and updates graph, we discard AS-sets from the data to avoid link ambiguity.
We filter private ASes~\cite{as-guidelines} because they create false links in the graph.
We then merge the 31 daily graphs of March~2004 into one graph for each BGP data source.

\begin{table*}[tbh]
    \centering
    \caption{{\bf Comparison of graphs built from different data sources.}
    The baseline graph~\mbox{$G_A$} is the BGP tables graph.
    Graph~\mbox{$G_B$} is the other graph listed in the first row.}
    \input{tables/comparison.tex}
    \label{table:comparison}
\end{table*}

We show the overlap statistics of our graphs in Table~\ref{table:comparison}.
This table uses the BGP-table graph as the baseline and compares it with
the BGP-updates graph in the first column.
Between the
two BGP-derived graphs, we note the similarity in the sets of
their constituent nodes and links. Given minor differences between node and link sets
of the BGP table- and update-derived topologies, we find the graph metric values
calculated for these two topologies to be nearly identical for all characteristics that we consider.
Therefore, in the rest of this study we present
characteristics of the static BGP table graph only and refer to it as the {\em BGP graph}.\footnote{Plots
and tables with metrics of the BGP-update graph included are available in~\cite{comp-anal}.}

{\bf Traceroute}~\cite{traceroute} captures the sequence of IP
hops along the forward path from the source to a given destination by sending
either UDP or ICMP probe packets to the destination.
CAIDA has developed a tool, {\em skitter}~\cite{skitter}, to collect continuous
traceroute-based Internet topology measurements. {\em skitter} maintains a target
destination list that comprises approximately one million IPv4 addresses.
CAIDA collects these addresses from various sources such as
existing destination lists, intermediate addresses in {\em skitter} traces, users
accessing CAIDA website. The goal is to find one responding IP address within each routable $/24$ segment, to provide
representative coverage of the routable IPv4 address space. The destination list is updated
once every 8 to 12 months to ensure the addresses stay current and to maximize
reachability. Skitter uses~25 monitors (traceroute sources), strategically
placed in the global Internet: 15 monitors in North America, 6 monitors in Europe,
3 monitors in Japan and 1 in New Zealand. Each monitor sends probe packets to destinations
in the target list and gathers the corresponding IP paths.

Using the core BGP tables provided by RouteViews, CAIDA maps the IP addresses in
the gathered IP paths to AS numbers, constructs the resulting AS-level topology graphs on a daily basis
and makes these graphs publicly available at~\cite{as-adjacencies}.
For this study, we start with daily graphs for each day
of March~2004, {\it i.e.},~31 daily graphs.
Mapping {\em skitter}-observed IP addresses to AS numbers involves potential distortion,
{\it e.g.,} due to multi-origin ASes, that is,
the same prefixes advertised by multiple ASes~\cite{MaReWaKa03},
AS-sets, and private ASes.
Both multi-origin ASes and AS-sets create ambiguous mappings between
IP addresses and ASes, hence we filter them from each graph. In addition, we filter
private ASes as they create false links. Unresolved IP hops in the traceroute data
give rise to indirect links~\cite{as-adjacencies}, which we also discard. The total discarded
and filtered links constitute approximately~5 percent of all links in the initial graph. We then merge all
the daily graphs to form one graph, which we call the {\em skitter graph}.

Comparing the skitter graph with the BGP graph (Table~\ref{table:comparison},
column 2 vs.\ baseline), we notice that there is exactly~1 node seen in the skitter but not
in the BGP graph. This node is AS2277 (Ecaunet).
Since we use BGP table dumps to map IP addresses to AS numbers
in constructing the skitter graph, we expect the number of nodes present in the skitter but
not in the BGP to be~$0$.
The one node difference occurs because different BGP table
dumps were used to construct the BGP table graph and to perform IP-to-AS mapping
in the skitter graph on the day when {\em skitter} observed
this IP address in its traces.

{\bf WHOIS}~\cite{irr} is a collection of databases with AS peering
information useful to network operators. These databases are
manually maintained with little requirements for timely updates of
registered information. Of the public WHOIS databases,
RIPE's WHOIS database contains the most reliable
current topological information, although it covers primarily
European Internet infrastructure~\cite{SiFa04,ChaGoJaSheWi04}.

We obtained the RIPE WHOIS database dump for April~07, 2004.
We are interested in the following types of records: {\tt
\begin{center}
\begin{tabular}{ll}
aut-num:&ASx\\
import:&from ASy\\
export:&to ASz
\end{tabular}
\end{center}}
This record indicates the presence of links between ASx-ASy and ASx-ASz.
We construct an AS-level graph (here after referred to as {\em WHOIS graph}) from these records and
exclude ASes that did not appear in the {\tt aut-num} lines. Such ASes are external to
the database and we cannot correctly estimate their topological
properties, e.g.,~node degree. We also filter private ASes.

Both Table~\ref{table:comparison} (column 3) and
the topology metrics we consider in Section~\ref{sec:characteristics} show
that the WHOIS topology differs significantly from the other two graphs. Thus,
the following question arises: Can we explain the difference by the fact that the WHOIS graph
contains only a part of the Internet, namely European ASes? To answer this
question we perform the following experiment. We consider the BGP tables
and WHOIS topologies narrowed to the set of nodes present both in BGP tables
and WHOIS,  {\it i.e.}, the 5,583 nodes present
in the intersection of BGP tables  and WHOIS graphs (Table~\ref{table:comparison}) and compute the
various topological characteristics for these reduced graphs. We then compare the
properties of the original BGP and WHOIS graphs to their reduced graphs respectively and find
that the reduced graphs preserve the full set of the properly normalized
topological properties of the original graphs. In other words, the reduced BGP graph,
consisting only of ASes found in the intersection of WHOIS and the original BGP graph,
has topological characteristics similar to the original BGP graph, while the reduced
WHOIS graph has characteristics similar to the original WHOIS graph.
Therefore, the differences between full BGP and WHOIS topologies are likely due to
dissimilar intrinsic properties of their originating data sources, and not due to
geographical biases in sampling the Internet.

Based on the very method of their construction, the three graphs in
this study reveal different sides of the actual Internet AS-level
topology. The skitter graph closely reflects the topology of actual
Internet traffic flows, {\it i.e.},~the data plane. The BGP graph reveals the
topology seen by the routing system, {\it i.e.},~the control plane. The BGP graph
does not reflect how traffic actually travels toward a destination network.
The WHOIS graph reflects the topology extracted from manually maintained
databases, {\it i.e.},~the management plane.

\subsection{Limitations and validity of our results}

All our data sources have some inaccuracies arising from their collection methodology.
Since {\em skitter} methodology relies on answers to ICMP requests,
ICMP filtering at intermediate hops adds some inaccuracy to the data. {\em skitter} also fails
to receive ICMP replies in the address blocks advertised by some small ASes.
The BGP graph depends on routing table exchanges, and not all peer ASes
advertise all their peering relationships; therefore the BGP graph tends to
miss these unadvertised links. Various misconfigurations, {\it e.g.,}
announcement of prefixes not owned by an AS, {\it etc.,} are
some of the other causes of errors with the BGP data. The manually maintained WHOIS
database is most likely to contain stale or inaccurate information~\cite{SiFa04}.
In fact, the WHOIS graph is likely to reflect unintentional or
even intentional over-reporting of peering relationships by some providers. There have been
reports about some ISPs entering inaccurate information in the WHOIS database
to increase their ``importance'' in the Internet hierarchy~\cite{SiFa04}.

We limit our data collection to a single month for obtaining the skitter and BGP graph.
If the topology of the Internet evolves with time, then the values of metrics
that we calculate might also change. While we believe that the interdependencies
between different metrics will hold for data gathered over various periods of
time and are not an artifact of the current Internet or our sampling period,
we leave this study to future work.

When processing each of our data sets to create the desired graph, we make choices
while dealing with ambiguities and errors in the raw data. One
example is the detection of ``false'' links created by route changes in
traceroute data. The processing we apply may potentially cause ambiguity in our final graphs.

While all three sources of topology data contain a number of sources of errors and
cannot be considered perfect representations of true AS-level interconnectivity,
the results of a number of recent studies indicate that the available data is a
reasonable approximation of AS topology.
The presence of global and strategically located vantage points for both BGP and skitter graphs as
well as the careful choice of destinations used by {\em skitter} lend credibility to traceroute-based
measurement studies. There have been some doubts about the validity of topologies obtained from
traceroute measurements. Specifically, Lakhina {\em et al.}~\cite{LaByCroXie03} numerically
explored sampling biases arising from traceroute measurements and found that such traceroute-sampled
graphs of the Internet yield insufficient evidence for characterizing the
actual underlying Internet topology. However, Dall'Asta
{\em et al.}~\cite{DaAlHaBaVaVe05} convincingly refute their conclusions
by showing that various traceroute exploration strategies provide sampled
distributions with enough signatures to statistically distinguish
between different topologies. The authors also argue that real mapping experiments
observe genuine features of the Internet, rather than artifacts.

%% file: tables/comparison.tex
\begin{tabular}{|l|c|c|c|}
\hline
&  BGP updates&  skitter&  WHOIS\\ \hline
 Number of nodes in both $G_A$ and $G_B$   ($|V_{A} \bigcap V_{B}|$) & 17,349 & 9,203 & 5,583 \\  
\hline
 Number of nodes in $G_A$ but not in $G_B$ ($|V_{A} \setminus V_{B}|$) & 97 & 8,243 & 11,863 \\  
\hline
 Number of nodes in $G_B$ but not in $G_A$ ($|V_{B} \setminus V_{A}|$) & 68 & 1 & 1,902 \\  
\hline
 Number of edges in both $G_A$ and $G_B$   ($|E_{A} \bigcap E_{B}|$) & 38,543 & 17,407 & 12,335 \\  
\hline
 Number of edges in $G_A$ but not in $G_B$ ($|E_{A} \setminus E_{B}|$) & 2,262 & 23,398 & 28,470 \\  
\hline
 Number of edges in $G_B$ but not in $G_A$ ($|E_{B} \setminus E_{A}|$) & 3,941 & 11,552 & 44,614 \\  
\hline
\end{tabular}

%% file: characteristics.tex
In this section, we quantitatively analyze differences between
the three graphs in terms of various topology metrics. We intentionally
do not introduce any new metrics: the set of characteristics we
discuss here encompasses most of the metrics discussed
in the networking literature before~\cite{TaGoJaShWi02,LiAlWiDo04,BuTo02,ZhoMo04}.
Relative to other studies, we analyze the broadest array
of network topology characteristics.

For each metric, we address the following points: 1)~metric definition;
2)~metric importance; and 3)~discussion on the metric values for the three measured topologies.
We present these results in the plots associated
with every metric and in the master Table~\ref{table:summary} containing all the scalar metric
values for all the three graphs.

We begin with simple metrics that characterize local connectivity in a network. We then move on to
metrics that describe global properties of the topology. These latter metrics play a vital role in
the performance of network protocols and applications.

\subsection{Average degree}

\textbf{\textit{Definition.}}
The two most basic graph properties are the
{\bf number of nodes}~$n$ (also referred to as {\bf graph size}) and the
{\bf number of links}~$m$. They define
the {\bf average node
degree}~\mbox{$\bar{k}=2m/n$}.

\textbf{\textit{Importance.}}
Average degree is the coarsest connectivity
characteristic of the topology. Networks with higher~$\bar{k}$ are ``better-connected'' on average
and, consequently, are likely to be more robust.
Detailed topology characterization based only on the average degree is rather limited,
since graphs with the same average node degree can have vastly different structures.

\textbf{\textit{Discussion.}}
The WHOIS graph has the smallest number of nodes,
but its average degree is almost three times larger than that of BGP, and
\mbox{$\sim 2.5$} times larger than that of skitter (Table~\ref{table:summary}).
In other words, WHOIS contains
substantially more links, both in the absolute~($m$) and relative~($\bar{k}$)
senses, than any other data source, although the credibility of these links is
lowest (cf.~Section~\ref{sec:data}). The chief reason for WHOIS graph's high average
degree lies in its measurement specifics: we have information from every node's perspective
in the database, while skitter and BGP graphs are obtained by sampling using tree-like explorations
of the Internet's ASes.

We also observe that the number of nodes in the BGP graph is almost
twice the number of nodes in skitter. This again can be explained by the
measurement techniques of the two data sources: {\em skitter} relies on responses
to ICMP requests sent to IP addresses on its target list of destinations and
it may not have any targets in the address blocks advertised by some small
ASes. As a result, {\em skitter} does not see these ASes.
The BGP routing tables however contain information about these ASes and thus these
nodes are observed in the
BGP graph. The extra ASes in the BGP dataset are mostly low-degree (cf.~Section~\ref{sec:degree-distr})
and therefore the BGP graph has a lower average degree than skitter.

Graphs ordered by increasing average degree~$\bar{k}$ are BGP,
skitter, WHOIS. We call this order the {\bf $\mathbf{\bar{k}}$-order}.

\subsection{Degree distribution}\label{sec:degree-distr}

\textbf{\textit{Definition.}}
Let~$n(k)$ be the number of nodes of degree~$k$
($k$-degree nodes). The {\bf node degree
distribution} is the probability that a randomly selected
node is $k$-degree: \mbox{$P(k)=n(k)/n$}. The degree distribution
contains more information about
connectivity in a given graph than the average degree, since given
a specific form of~$P(k)$ we can always restore the average degree by
\mbox{$\bar{k} = \sum_{k=1}^{k_{max}} kP(k)$},
where~$k_{max}$ is the {\bf maximum node degree} in the graph. If the
degree distribution in a graph of size~$n$ is a power law, \mbox{$P(k) \sim k^{-\gamma}$},
where~$\gamma$ is a positive {\bf exponent}, then~$P(k)$ has a natural
cut-off at the {\bf power-law maximum degree}~\cite{DorMen-book03}:
$k_{max}^{PL} = n^{1/(\gamma-1)}$.

\textbf{\textit{Importance.}}
The degree distribution is the most frequently used topology characteristic.
The observation~\cite{FaFaFa99} that the Internet's degree distribution follows
a power law had significant impact on network topology research:
Internet models before~\cite{FaFaFa99} failed to exhibit power
laws. Researchers also widely believed that an organized hierarchy existed
among the ASes in the Internet. However, the
authors of~\cite{TaGoJaShWi02} showed that
topologies derived from structural generators that incorporated hierarchies of AS tiers did not have much
in common with topologies obtained from real observed data. The smooth power law degree distribution
indicates that there are no organized
tiers among ASes. The power law distribution also implies substantial variability associated
with degrees of individual nodes.

\begin{figure*}[tbh]
    \centerline{
        \subfigure[PDF]
        {\includegraphics[width=2.2in]{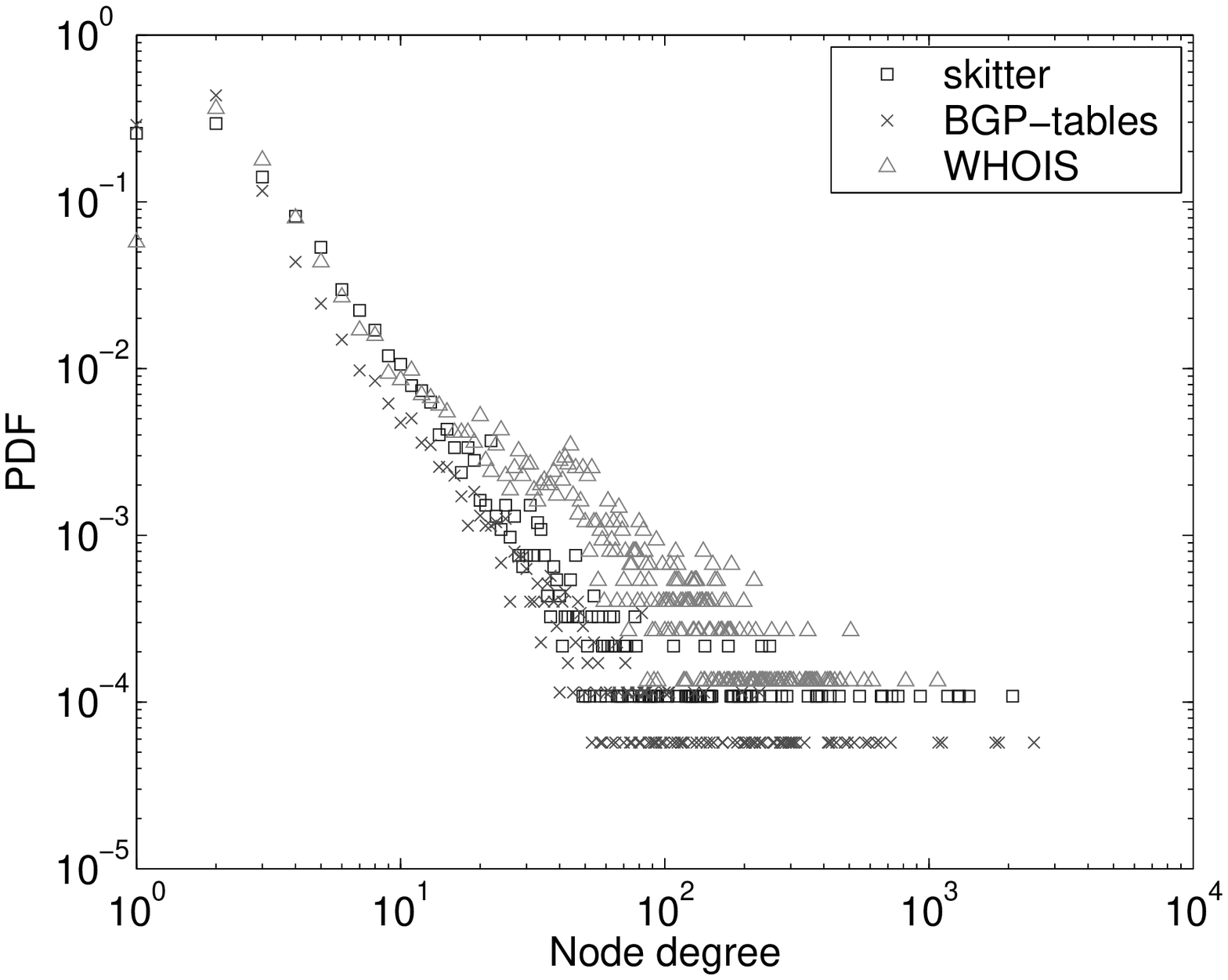}
        \label{fig:pk-pdf}}
        \hfill
        \subfigure[CCDF]
        {\includegraphics[width=2.2in]{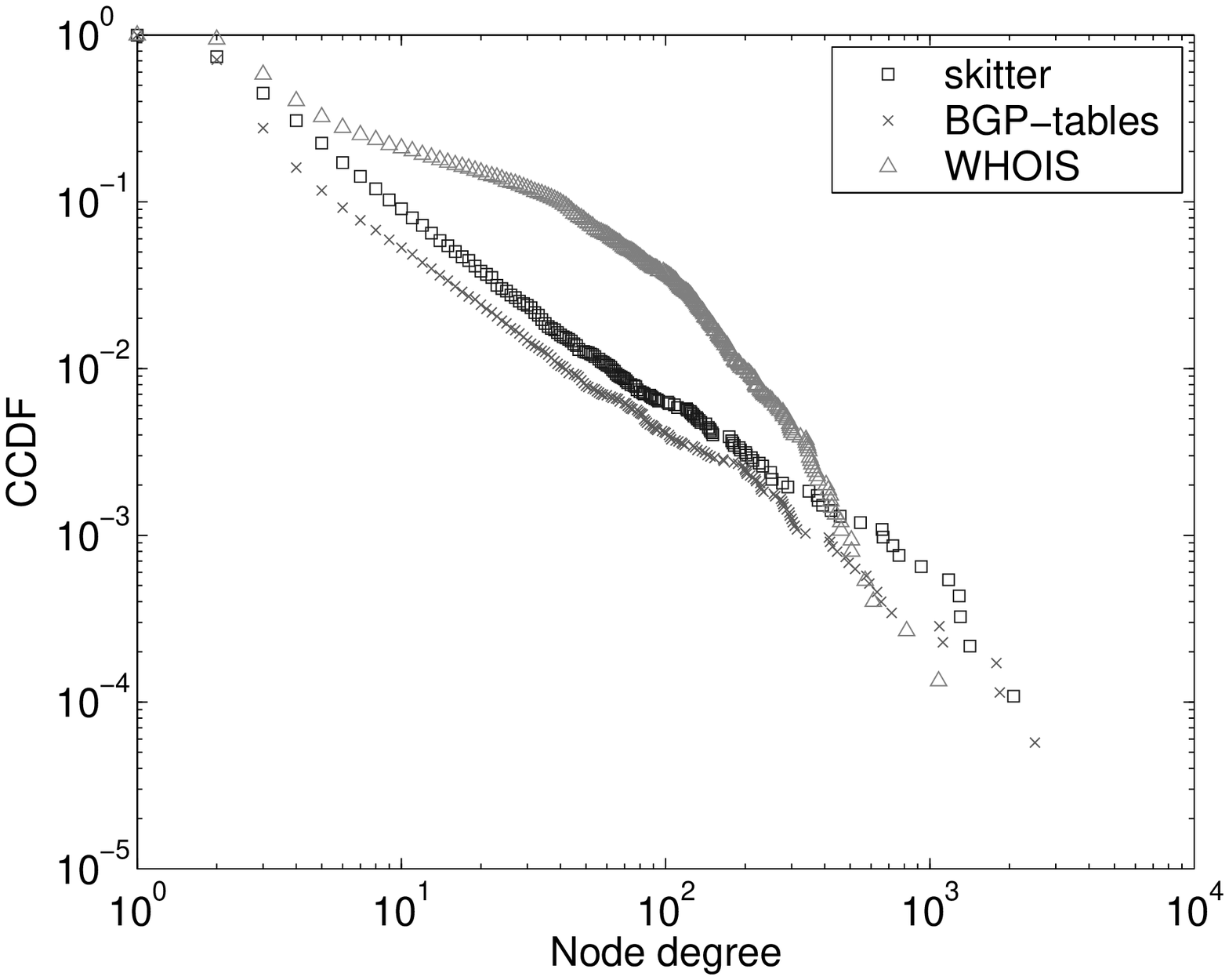}
        \label{fig:pk-ccdf}}
    \subfigure[PDFs of skitter vs.\ BGP differences]
        {\includegraphics[width=2.5in]{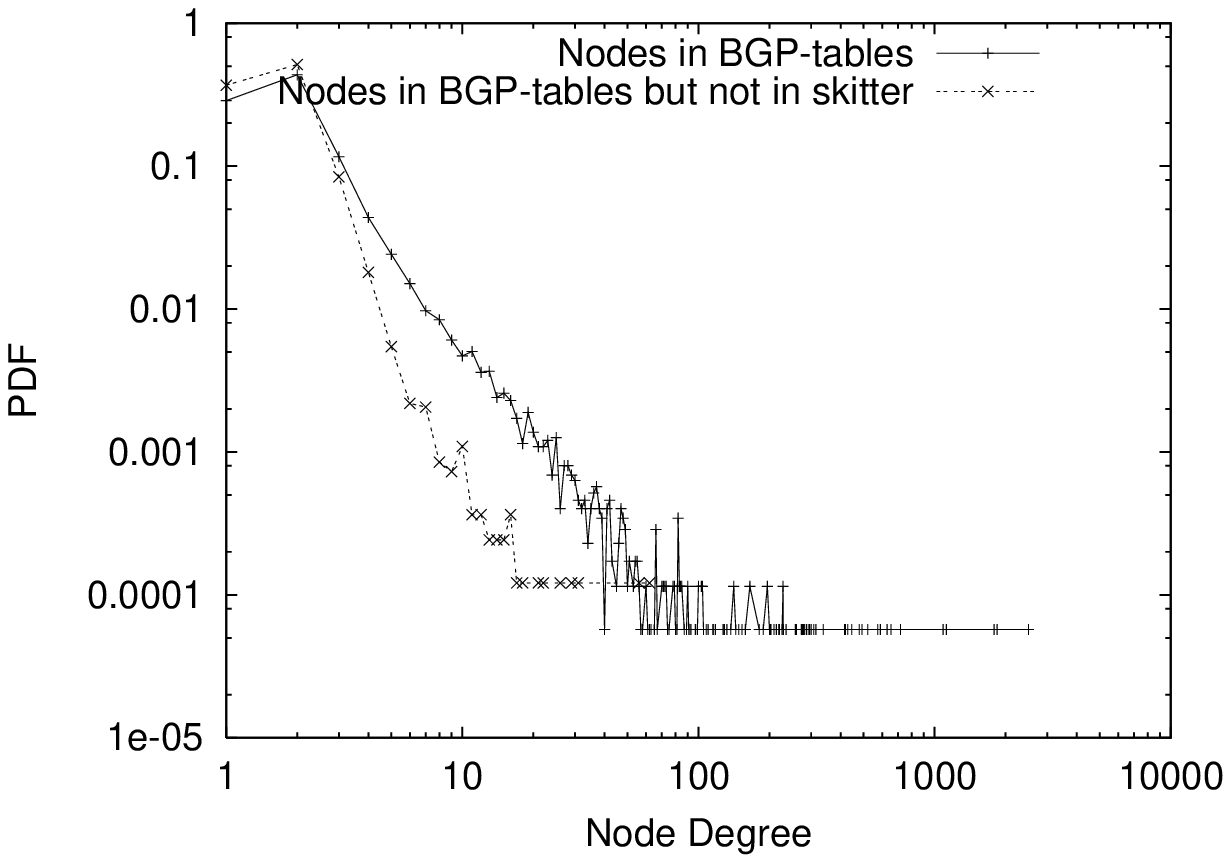}
        \label{fig:sk_vs_bgp:pk}}
    }
    \caption{\footnotesize \bf Node degree distributions~$\mathbf{P(k)}$.
}
    \label{fig:pk}
\end{figure*}

\textbf{\textit{Discussion.}}
As expected, the degree distribution PDFs and CCDFs
in Figure~\ref{fig:pk} are in the
$\bar{k}$-order (BGP $<$ skitter $<$ WHOIS) for a wide range of
node degrees.

Comparing the observed maximum node degrees~$k_{max}$ with those predicted
by the power law~$k_{max}^{PL}$ in Table~\ref{table:summary}, we conclude
that skitter is closest to power law. The power-law approximation for the BGP
graph is less accurate. The WHOIS graph has an excess of medium-degree nodes
and its node degree distribution does not follow a power law at all. It is not
surprising then that augmenting the BGP graph with WHOIS links breaks the power
law characteristics of the BGP graph~\cite{WilRevisited,ChaGoJaSheWi04}.

Note that there are fewer 1-degree nodes than 2-degree nodes in all the graphs
(Figure~\ref{fig:pk-pdf}). This effect is due to
the AS number assignment policies~\cite{as-guidelines}
allowing a customer to have an AS number only if it has multiple providers.
If these policies were strictly enforced and if there were no measurement
inaccuracies, then the minimum observed AS degree would be~2.

CCDFs of skitter and BGP graphs look similar (Figure \ref{fig:pk-ccdf}),
but Table~\ref{table:comparison} shows significant differences between the two
graphs in terms of (non-)intersecting nodes and links. We seek to answer the
question of where, topologically, these nodes and links are located.
Calculating the degree distribution of nodes present only in the BGP
graph (Figure~\ref{fig:sk_vs_bgp:pk}), we detect a skew toward low-degree nodes.
The average degree of the nodes that are present only in BGP graphs, but
not in skitter, is~$1.86$. {\em skitter}'s target list of
destinations to probe does not contain IP addresses that respond in the address
blocks advertised by these small ASes. As a result, the skitter graph misses
them. Most links present only in BGP, but not in skitter, are links between
low-degree ASes (see~\cite{comp-anal} for details). The majority of such links
connect the low-degree ASes present only in BGP to their secondary (backup)
low-degree providers, while their primary providers are of high degrees.
Even if {\em skitter} detects a low-degree AS having such a small backup provider,
the tool is still unlikely to detect the backup link since its traceroutes follow
the primary path via the large provider.

\subsection{Joint degree distribution}\label{sec:joint-degree}

While the node degree distribution tells us how many nodes of a given
degree are in the network, it fails to provide information on the
interconnection between these nodes: given~$P(k)$, we still do not know anything about
the structure of the neighborhood of the average node of a given degree.
The joint degree distribution fills this
gap by providing information about 1-hop neighborhoods around a node.

\textbf{\textit{Definition.}}
Let~$m(k_1,k_2)$ be the total number of edges connecting nodes of
degrees~$k_1$ and~$k_2$. The {\bf joint degree
distribution} (JDD), or the {\bf node degree correlation} matrix,
is the probability that a randomly selected
edge connects $k_1$- and $k_2$-degree nodes:
$P(k_1,k_2) = \mu(k_1,k_2) \times m(k_1,k_2)/(2m)$, where~$\mu(k_1,k_2)$
is~$1$ if~$k_1=k_2$ and~$2$ otherwise.
Note that $P(k_1,k_2)$ is different from the conditional
probability
$P(k_2|k_1)=(\bar{k} P(k_1,k_2) )/(k_1 P(k_1) )$ that a
given $k_1$-degree node is connected to a $k_2$-degree node.
The JDD contains more information about
the connectivity in a graph than the degree distribution, since given a
specific form of~$P(k_1,k_2)$ we can always restore both the degree
distribution~$P(k)$ and average degree~$\bar{k}$ by expressions
in~\cite{DorMen-book03}.
A summary statistic of JDD
is the {\bf the average neighbor
connectivity}~\mbox{$k_{nn}(k) = \sum_{k'=1}^{k_{max}} k' P(k'|k)$}.
It is simply the average neighbor degree of the average $k$-degree node.
It shows whether ASes of a given degree preferentially
connect to high- or low-degree ASes.
In a full mesh graph, $k_{nn}(k)$ reaches its maximal possible value, $n-1$.
Therefore, for uniform graph comparison we plot normalized values \mbox{$k_{nn}(k)/(n-1)$}.
We can further summarize the JDD by a single scalar called
{\bf assortativity coefficient}~$r$~\cite{newman02,dorogovtsev03},
$r \sim \sum_{k_1,k_2=1}^{k_{max}} k_1 k_2 (P(k_1,k_2)- k_1 k_2 P(k_1) P(k_2)/ {\bar{k}}^2)$.

\textbf{\textit{Importance.}}
The assortativity coefficient~$r$, \mbox{$-1 \leq r \leq 1$}, has direct practical implications.
{\bf Disassortative} networks with \mbox{$r<0$} have an excess of {\bf radial} links, that is, links
connecting nodes of dissimilar degrees. Such networks are vulnerable to both random
failures and targeted attacks. On a positive note, vertex covers in disassortative graphs are smaller,
which is important for applications such as traffic monitoring~\cite{BreChaGaRaSi01} and prevention
of DoS attack~\cite{PaLe01}. The opposite properties apply to {\bf assortative} networks with
\mbox{$r>0$} that have an excess of {\bf tangential} links, that is, links connecting nodes of similar
degrees.\footnote{The semantics behind the terms ``radial'' and
``tangential'' comes from the commonly used technique in visualization of the large-scale
Internet topologies~\cite{skitter-poster,AlDaBaVe04,TaPaSiFa01}: high-degree
nodes populate the center of a circle, while low-degree nodes are
close to the circumference. Links connecting high-degree nodes to low-degree nodes
are indeed radial then.}

In contrast to the widely studied degree distribution, the network community has
only recently started recognizing the importance of JDD~\cite{WiJa02,JaRoTo04}.
In the most prominent recent example~\cite{LiAlWiDo04} Li {\em et al.}~define
{\em likelihood} and make this metric central for their argument.
They propose to use
likelihood, which is directly related to the assortativity coefficient, as a
measure of randomness to differentiate between multiple
graphs with the same degree distribution. Such a measure is important
for evaluating the amount of order, e.g., engineering design constraints,
present in a given topology. A topology with low likelihood is not random;
it results from some sophisticated evolution processes involving specific
design purposes.

\textbf{\textit{Discussion.}}
All the three Internet
graphs built from our data sources are disassortative (\mbox{$r<0$}) as seen in
Table~\ref{table:summary}. We call the order of graphs with decreasing
assortativity coefficient~$r$---WHOIS, BGP, skitter---the
{\bf $\mathbf{r}$-order}.

We can explain the $r$-order in terms of differing topology measurement methodologies.
First, we notice that both skitter and BGP graphs are results of {\em tree-like\/} explorations of the network topology,
meaning that we can roughly approximate these graphs by a union of spanning trees
rooted at, respectively, {\em skitter} monitors or BGP data collection points. As such,
both these methods are likely to discover more radial links connecting numerous low-degree
nodes, {\it i.e.,} small ASes, to high-degree nodes,
{\it i.e.,} large ISP ASes, where the monitors are located.
At the same time, these measurements fail to detect some
tangential links interconnecting medium-to-low degree
nodes since many of these links belong to none
of the spanning trees rooted at the vantage points in the core.
In contrast, WHOIS data contains abundant medium-degree tangential
links because it relies on operators to report {\em all\/} the links
attached to a given AS, {\it i.e.,} a source of a WHOIS record.
This excess of tangential links in WHOIS is thus responsible for its much
higher assortativity.
Second, we explain that the BGP graph has a slightly higher assortativity
than the skitter graph. As discussed in Section~\ref{sec:degree-distr},
the BGP graph contains the tangential links between low-degree nodes that
traceroute probes of {\em skitter} miss since these links are typically the
backup links to smaller secondary providers, while {\em skitter}'s ICMP packets
tend to follow the primary paths to larger primary providers. This small excess
of tangential links is responsible for a slightly higher assortativity of the
BGP graph compared to skitter.

\begin{figure*}[tbh]
  \begin{minipage}[t]{2.2in}
      \centerline{
          \includegraphics[width=2.2in]{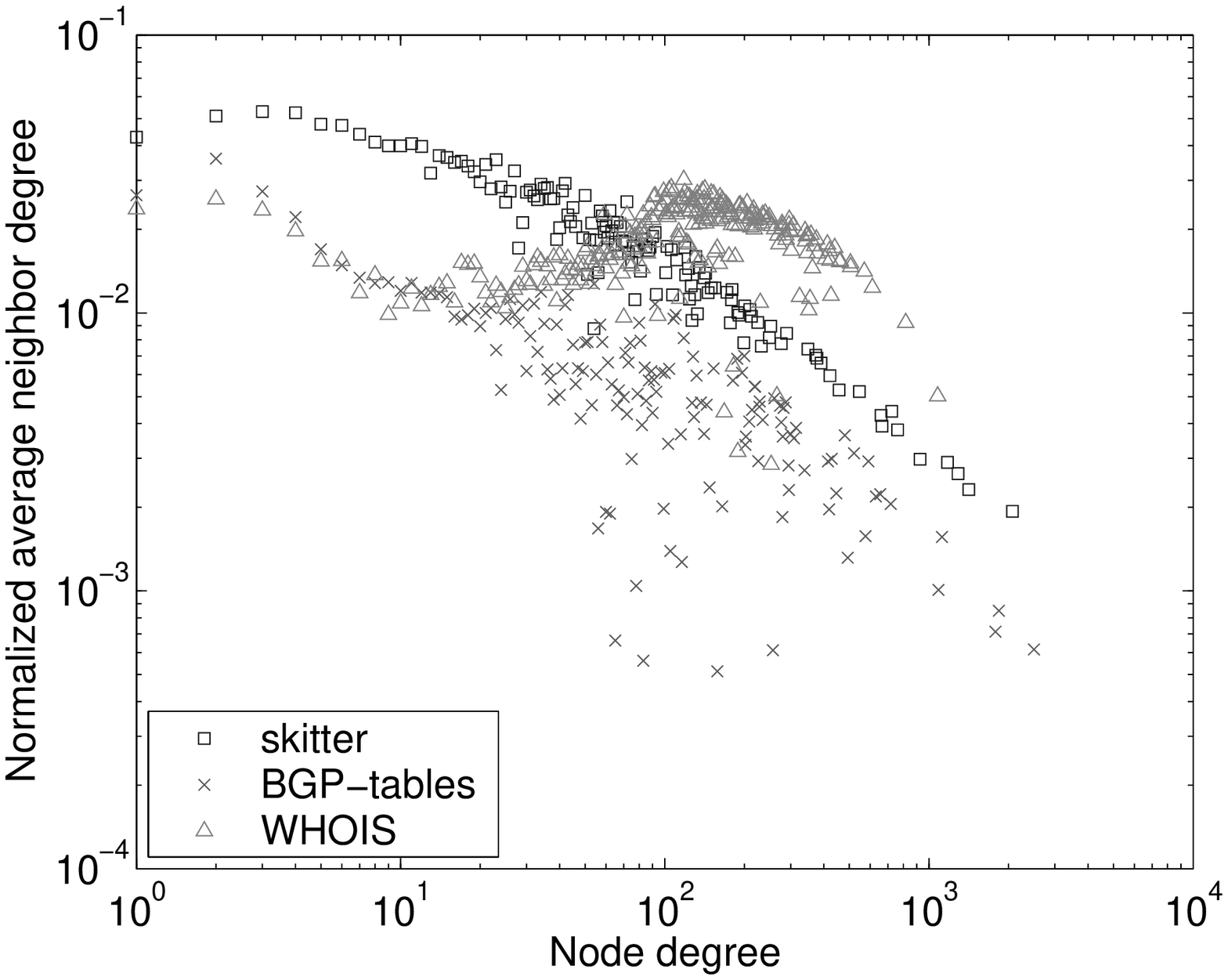}
      }
      \caption{\footnotesize \bf Normalized average neighbor
        connectivity~$\mathbf{k_{nn}(k)/(n-1)}$.
      }
      \label{fig:knnk}
  \end{minipage}
  \hfill
  \begin{minipage}[t]{2.2in}
      \centerline{
          \includegraphics[width=2.2in]{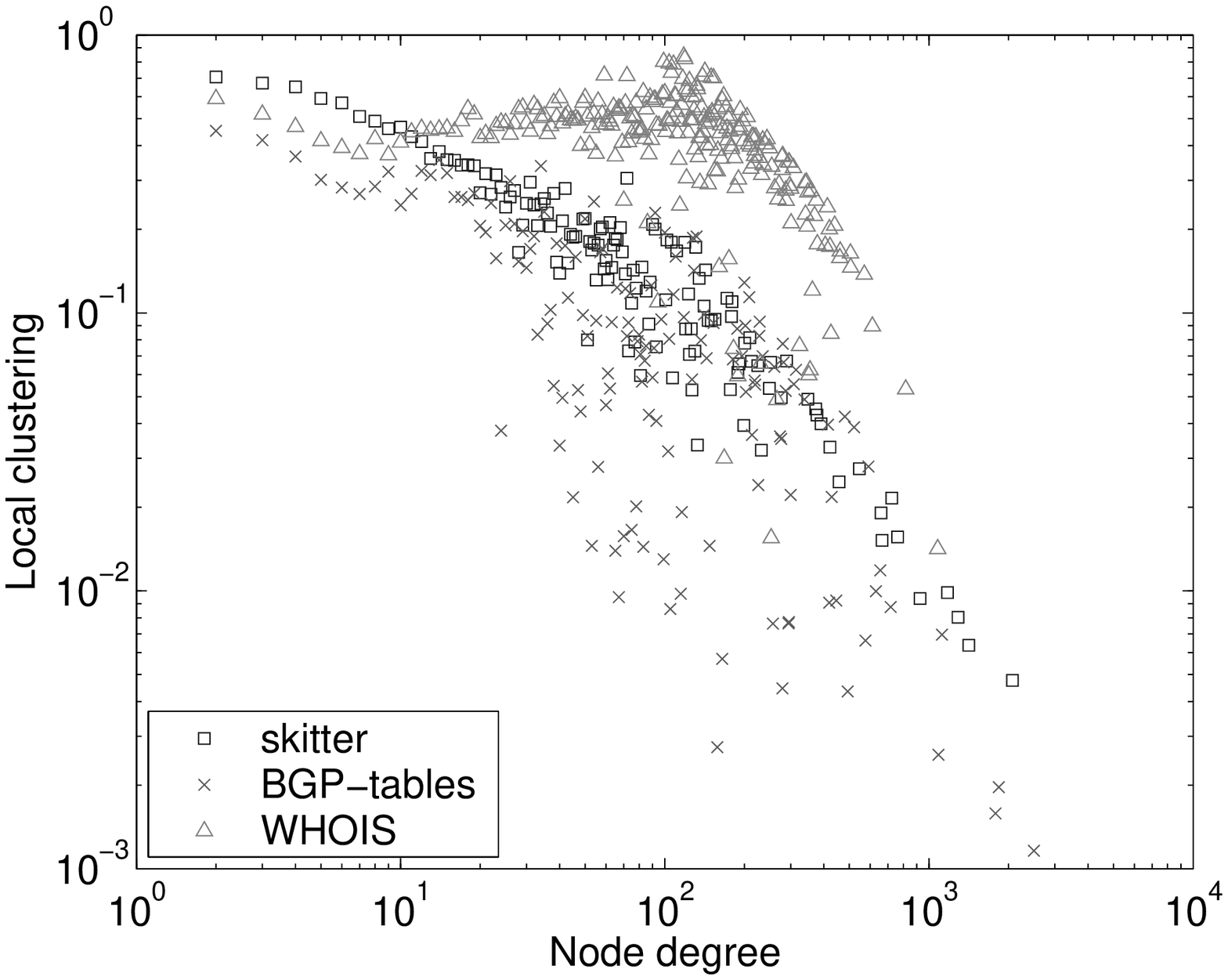}
      }
      \caption{\footnotesize \bf Local clustering~$\mathbf{C(k)}$.
      }
      \label{fig:ck}
  \end{minipage}
\hfill
  \begin{minipage}[t]{2.2in}
      \centerline{
    \includegraphics[width=2.2in]{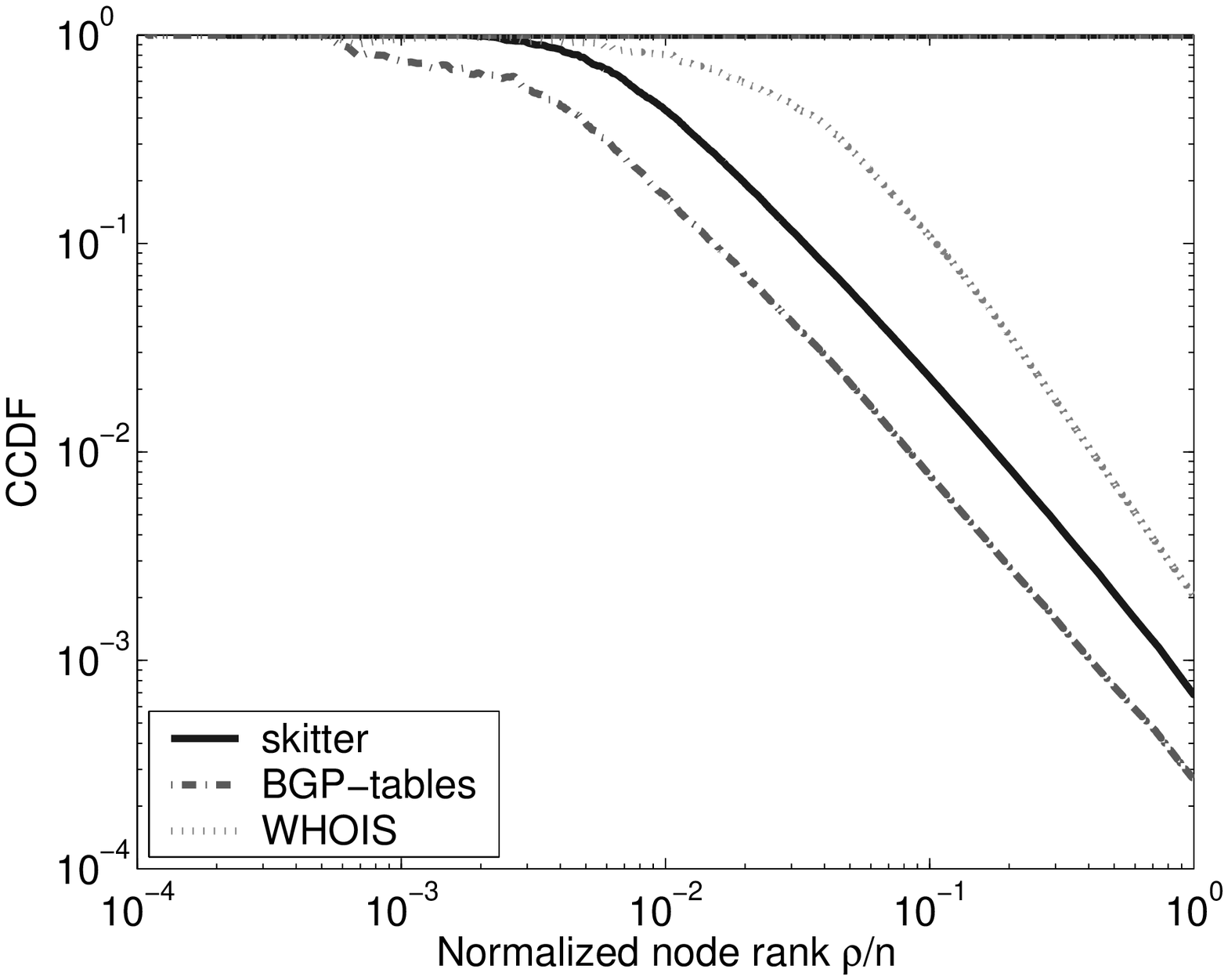}
    }
      \caption{\footnotesize \bf Rich club connectivity~$\mathbf{\phi(\rho/n)}$.}
      \label{fig:rich_club}
 \end{minipage}
\end{figure*}

The interplay between $\bar{k}$- and $r$-orders underlies Figure~\ref{fig:knnk},
where we plot the average neighbor connectivity functions
for the three graphs.
Skitter has the largest excess of radial links that
connect low-degree nodes (customers ASes) to high-degree nodes
(large provider ASes). The highest relative number of radial links
is responsible for skitter's
highest average degree of the neighbors of low-degree nodes:
in Figure~\ref{fig:knnk}, skitter is at the top in the area
of low degrees, while BGP is below and WHOIS is at the bottom
($r$-order). On the other hand, the greatest
proportion of tangential links
between ASes of similar degrees in the
 WHOIS graph contributes to connectivity of
neighbors of high-degree nodes; therefore the WHOIS graph is at the top
for high-degree nodes ($\bar{k}$-order).

Note that in the case of skitter and BGP, \mbox{$k_{nn}(k)$}
can be approximated by a power law with the corresponding
exponents $\gamma_{nn}$ in Table~\ref{table:summary}.

\subsection{Clustering}\label{sec:clustering}

While JDD contains information about the degrees of neighbors for
the average $k$-degree node, it does not tell us how
these neighbors interconnect. Clustering partially satisfies this  need
by providing a measure of how close a node's neighbors are to forming a clique.

\textbf{\textit{Definition.}}
Let~\mbox{$\bar{m}_{nn}(k)$} be the average number of
links between the neighbors of $k$-degree nodes.
{\bf Local clustering} is the ratio of this number to
the maximum possible number of such
links: \mbox{$C(k)= \bar{m}_{nn}(k)/{k \choose 2}$}. If two neighbors of a node are connected, then these three nodes
together form a triangle (3-cycle). Therefore, by definition, local clustering
is the average number of 3-cycles involving $k$-degree nodes. The two summary
statistics associated with local clustering are {\bf mean local clustering}~$C_{mean}=\sum C(k)P(k)$,
which is the average value of~$C(k)$, and the {\bf clustering coefficient}~$C_{coeff}$,
which is the percentage of 3-cycles among all connected node triplets in the
entire graph (for the exact definition, see~\cite{BoRi02}).

\textbf{\textit{Importance.}} Clustering expresses local robustness in
the graph and thus has practical implications: the higher the local
clustering of a node, the more interconnected are its neighbors, thus increasing the path diversity
locally around the node.
Networks with strong clustering are likely to be
chordal or of low
chordality,\footnote{{\em Chordality\/} of a graph is the length of
the longest cycle without chords. A graph is called {\em chordal\/}
if its chordality is~3.} which makes certain
routing strategies perform better~\cite{fraigniaud05}. One can also use
clustering as a litmus test for verifying the accuracy of a topology
model or generator~\cite{BuTo02}.

\textbf{\textit{Discussion.}}
We first observe that the clustering average values~$C_{mean}$
in Table~\ref{table:summary} are in the
$\bar{k}$-order, which is expected: clustering increases with increase in
number of links.
The values of~$C_{mean}$ are almost equal for
skitter and WHOIS, but the clustering coefficient~$C_{coeff}$ is
15 times larger for WHOIS than for skitter. As shown in~\cite{SoVa04},
orders of magnitude difference between~$C_{mean}$ and~$C_{coeff}$ is intrinsic
to highly disassortative networks and is a consequence of
strong degree correlations~(JDD) necessarily present in such
networks.

Similar to~$k_{nn}(k)$,
the interplay between $\bar{k}$- and $r$-orders explains Figure~\ref{fig:ck},
where we plot
local clustering as a function of node degree~$C(k)$. Skitter's clustering is the highest
amongst the three graphs for low-degree nodes because this graph is most disassortative. The links
adjacent to low-degree nodes are most likely to lead to high-degree nodes, the latter being
interconnected with a high probability. The WHOIS graph exhibits the highest
values for clustering for high-degree nodes since this graph has the highest average connectivity
(largest~$\bar{k}$). The neighbors of high-degree nodes are interconnected to
a greater extent, resulting in higher clustering for such nodes.

Similar to~$k_{nn}(k)$, $C(k)$ also can be
approximated by a power law for skitter and BGP graphs (exponents~$\gamma_C$
in Table~\ref{table:summary}).

Strong correlations in JDD play a major part for the presence of
non-trivial clustering observed in many networks~\cite{SoVa04}.
The interplay between $\bar{k}$- and $r$-orders explains
the overall similarity between degree correlations and
clustering, in general, and similarity between $k_{nn}(k)$ and~$C(k)$,
in particular.

\subsection{Rich club connectivity}

\textbf{\textit{Definition.}}
Let~\mbox{$\rho = 1 \ldots n$} be the first~$\rho$ nodes ordered by
their non-increasing degrees in a graph of size~$n$. {\bf Rich club
connectivity}~(RCC)~$\phi(\rho/n)$ is the ratio of the number of links in
the subgraph
induced by the~$\rho$ largest-degree nodes to the maximum possible number of such
links~$\rho \choose 2$. In other words, the RCC is a measure of how
close $\rho$-induced subgraphs are to cliques.

\textbf{\textit{Importance.}}
The Positive Feedback Preference~(PFP) model by Zhou and Mondragon~\cite{ZhoMo04}
has successfully reproduced a wide spectrum of metrics
of their measured AS-level topology by trying to explicitly capture only
the following three characteristics: (i)~the exact form of the node
degree distribution; (ii)~the maximum node degree; and
(iii)~RCC.
One can show that networks with the same
JDDs have the same RCC. The converse is not true, but given a specific
form of RCC, one can fully describe all possible JDDs that would yield
the specified RCC.

\textbf{\textit{Discussion.}}
As expected, the values of~\mbox{$\phi(\rho/n)$} in Figure~\ref{fig:rich_club}
are in the $\bar{k}$-order with WHOIS at the top: more links result in denser
cliques. RCC exhibits clean power laws for all three graphs in the area of
medium and large~\mbox{$\rho/n$}. The values of the power-law exponents~$\gamma_{rc}$ in Table~\ref{table:summary}
result from fitting~$\phi(\rho/n)$ with power laws
for~90\% of the nodes, \mbox{$0.1 \leq \rho/n \leq 1$}.

\subsection{Distance}\label{sec:characteristics:distance}

\textbf{\textit{Definition.}}
The shortest path length distribution or
simply the {\bf distance
distribution}~\mbox{$d(x)$} is the probability that a random pair of nodes are
at a distance~$x$ hops from each other. Two basic summary statistics
associated with the distance distribution of a graph are {\bf average
distance}~$\bar{d}$ and the {\bf standard deviation}~$\sigma$. We call the
latter the {\em distance distribution width} since distance distributions in
 Internet graphs (and in many other networks) have a characteristic
Gaussian-like shape.

\begin{figure*}[tbh]
  \begin{minipage}[t]{2.2in}
      \centerline{
          \includegraphics[width=2.25in]{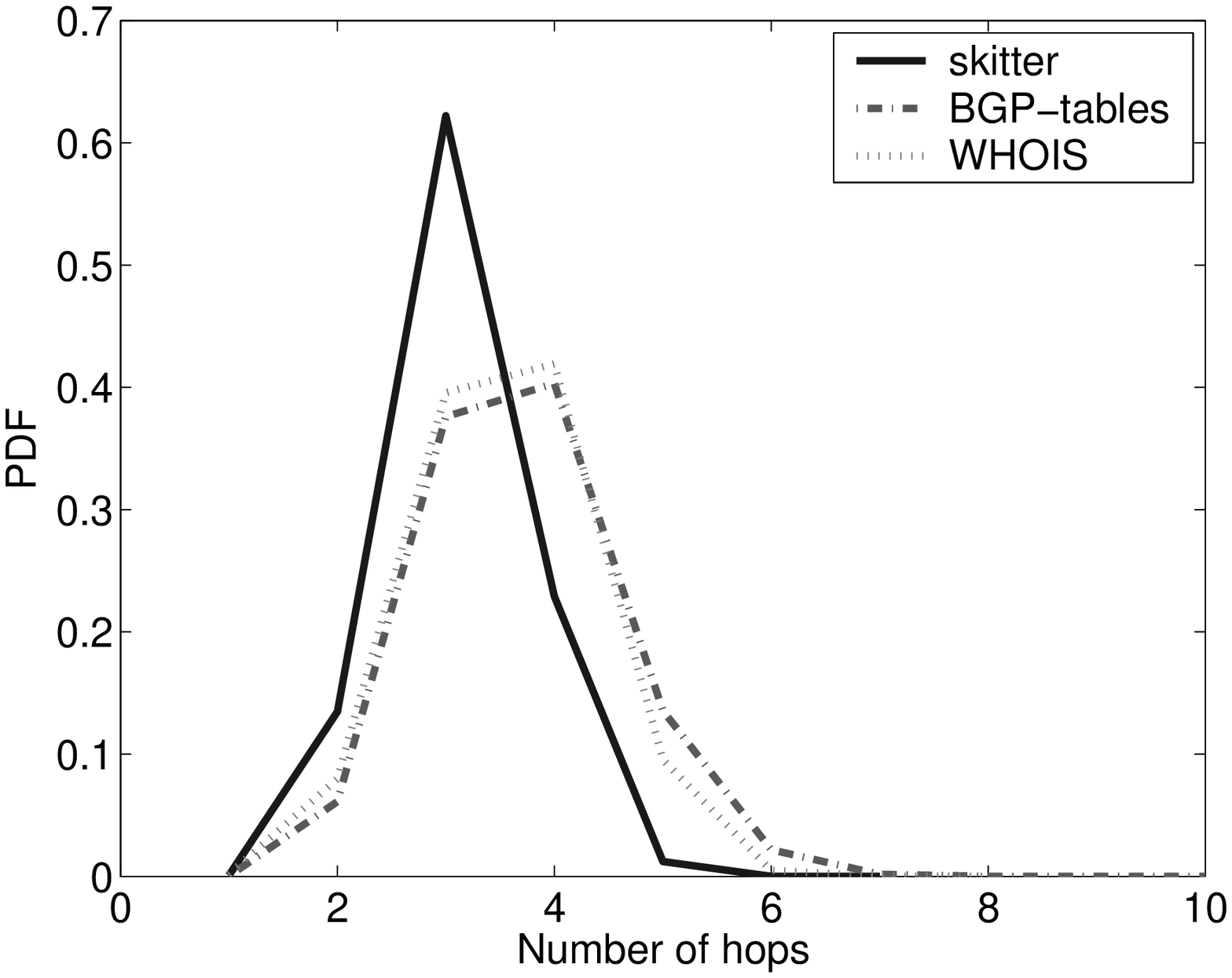}
      }
      \caption{\footnotesize \bf Distance~$\mathbf{d(x)}$ distribution.
    }
      \label{fig:distance-pdf}
  \end{minipage}
  \hfill
  \begin{minipage}[t]{2.2in}
      \centerline{
          \includegraphics[width=2.25in]{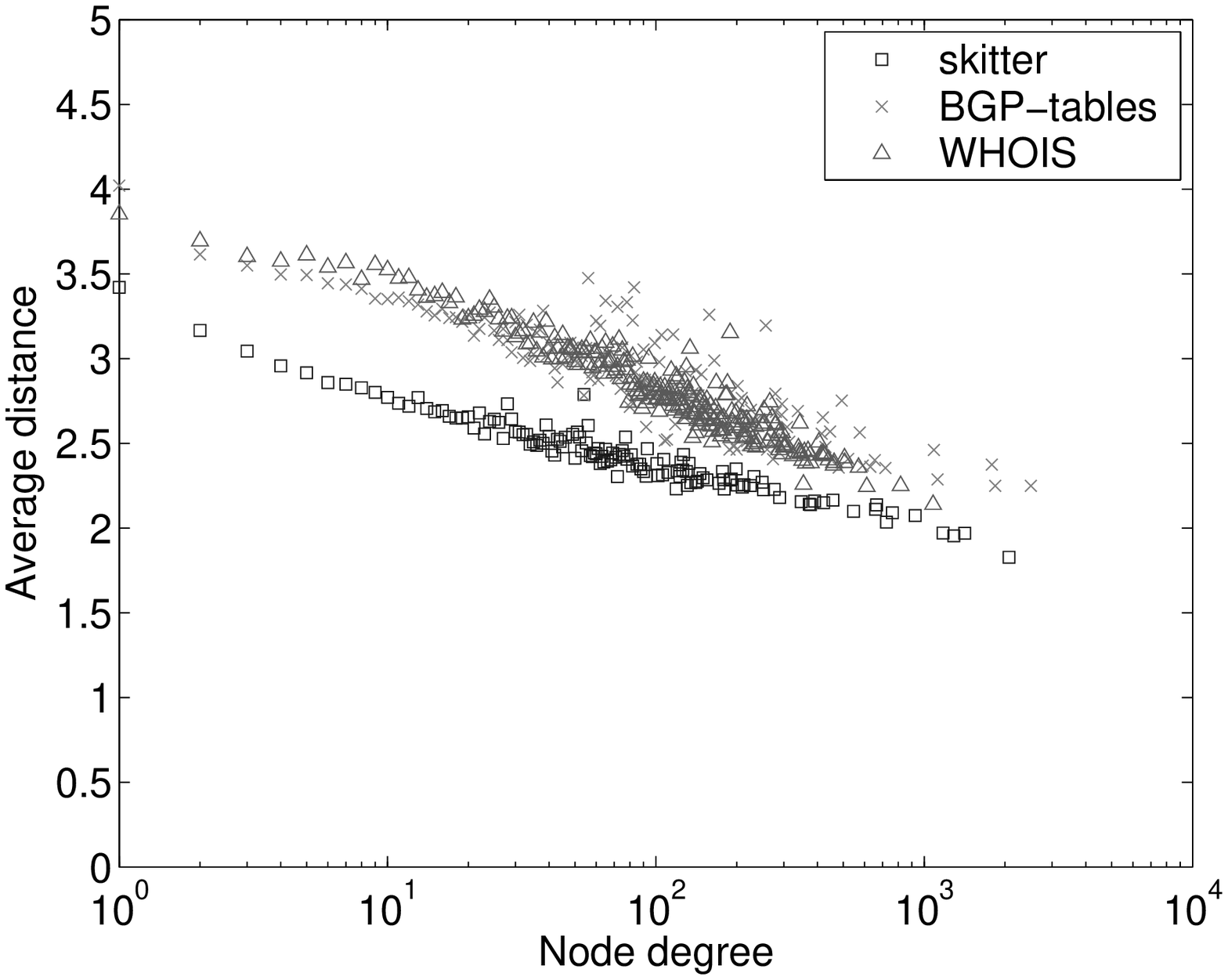}
      }
      \caption{\footnotesize \bf Average distance from $k$-degree nodes~$\mathbf{d(k)}$.}
      \label{fig:distance-k}
  \end{minipage}
  \hfill
 \begin{minipage}[t]{2.2in}
      \centerline{
          \includegraphics[width=2.25in]{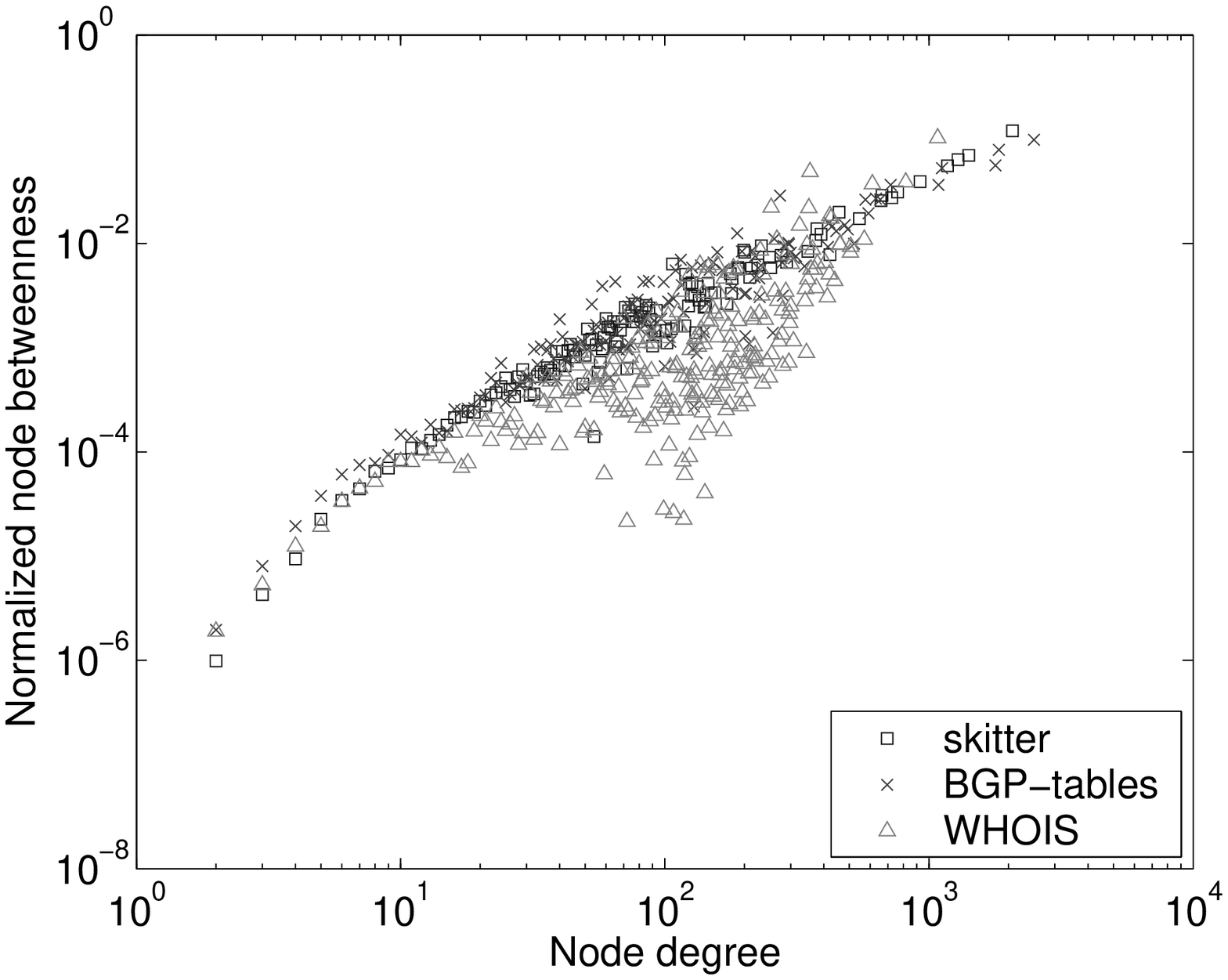}
      }
      \caption{\footnotesize \bf Normalized node betweenness~$\mathbf{B(k)/n/(n-1)}$.
}
      \label{fig:betweenness}
  \end{minipage}
  \hfill
\end{figure*}

\textbf{\textit{Importance.}}
Distance distribution is important for many applications,
the most prominent
being routing. A distance-based locality-sensitive approach~\cite{peleg01}
is the root of most modern routing algorithms. As shown
in~\cite{KrFaYa04}, performance parameters of these algorithms depend mostly on
the distance distribution. In particular, short average
distance and narrow distance distribution width break the efficiency of
traditional hierarchical routing. They are among the root causes
of interdomain routing scalability issues in the Internet today.

Distance distribution also plays a vital role in robustness of the network to worms. Worms
can quickly contaminate a network that has
small distances between nodes. Topology models that accurately reproduce
observed distance distributions will benefit researchers
developing techniques to quarantine the network from worms~\cite{ShMo04}.

We note that {\em expansion\/}, identified in~\cite{TaGoJaShWi02}
as a critical metric for topology comparison analysis, is a
renormalized version
of the distance distribution: it is the product of the
distance distribution and the graph size~$n$.

\textbf{\textit{Discussion.}}
Although the distance distribution is a global topology characteristic,
we can explain Figure~\ref{fig:distance-pdf} by the interplay between our
local connectivity characteristics: the
$\bar{k}$- and $r$-orders. First, we note
that the skitter graph stands out in Figure~\ref{fig:distance-pdf} as it has the
smallest average distance and the smallest distribution width
(Table~\ref{table:summary}). This result appears unexpected at first since the
skitter graph has more nodes than the WHOIS graph and only about half the
links. One would expect a denser graph (WHOIS) to have a lower average distance
since adding links to a graph can only {\em decrease\/} the
average distance in it. Surprisingly, the average distance of the most richly
connected (highest~$\bar{k}$) WHOIS graph is not the lowest. This result can be explained using
the $r$-order.
Indeed, a more disassortative graph has a greater proportion of radial links,
shortening the distance from the fringe to the core.\footnote{We
use terms {\em fringe} and {\em core} to mean ``zones'' in the graph with low-
and high-degree nodes respectively, cf.~\cite{TaPaSiFa01}.}
The skitter graph has the right balance between the
relative number of links~$\bar{k}$ and their radiality~$r$,
that minimizes the average distance. Compared to skitter, the BGP graph has larger
distance because it is sparser
(lower~$\bar{k}$), and the WHOIS graph has larger distance because it is more
assortative (higher~$r$).

Another observation is that for all three graphs, including WHOIS,
the average distance as a function of node degree exhibits relatively stable
power laws in the full range of node degrees (Figure~\ref{fig:distance-k}),
with exponents given in Table~\ref{table:summary}.

\subsection{Betweenness}\label{sec:characteristics:betweenness}

Although the average distance is a good node centrality measure---intuitively,
nodes with smaller
average distances are closer to the graph ``center,''---the most
commonly used measure of centrality is betweenness. It is applicable not
only to nodes, but also to links.

\textbf{\textit{Definition.}}
Betweenness measures the number of shortest paths
passing through a node or link and, thus, estimates the potential traffic load
on this node/link assuming uniformly distributed traffic following shortest
paths. Let~$\sigma_{ij}$ be the number
of shortest paths between nodes~$i$ and~$j$ and let~$l$ be either a node
{\em or\/} link. Let~$\sigma_{ij}(l)$ be the number of shortest
paths between~$i$ and~$j$ going through node (or link)~$l$. Its
{\bf betweenness} is $B_l = \sum_{ij}\sigma_{ij}(l)/\sigma_{ij}$.
The maximum possible value for node and link betweenness is \mbox{$n(n-1)$}~\cite{DaAlHaBaVaVe05},
therefore in order to compare betweenness in graphs of different sizes, we
normalize it by \mbox{$n(n-1)$}.

\begin{figure*}[tbh]
    \centerline{
        \subfigure[skitter]
        {\includegraphics[width=2in]{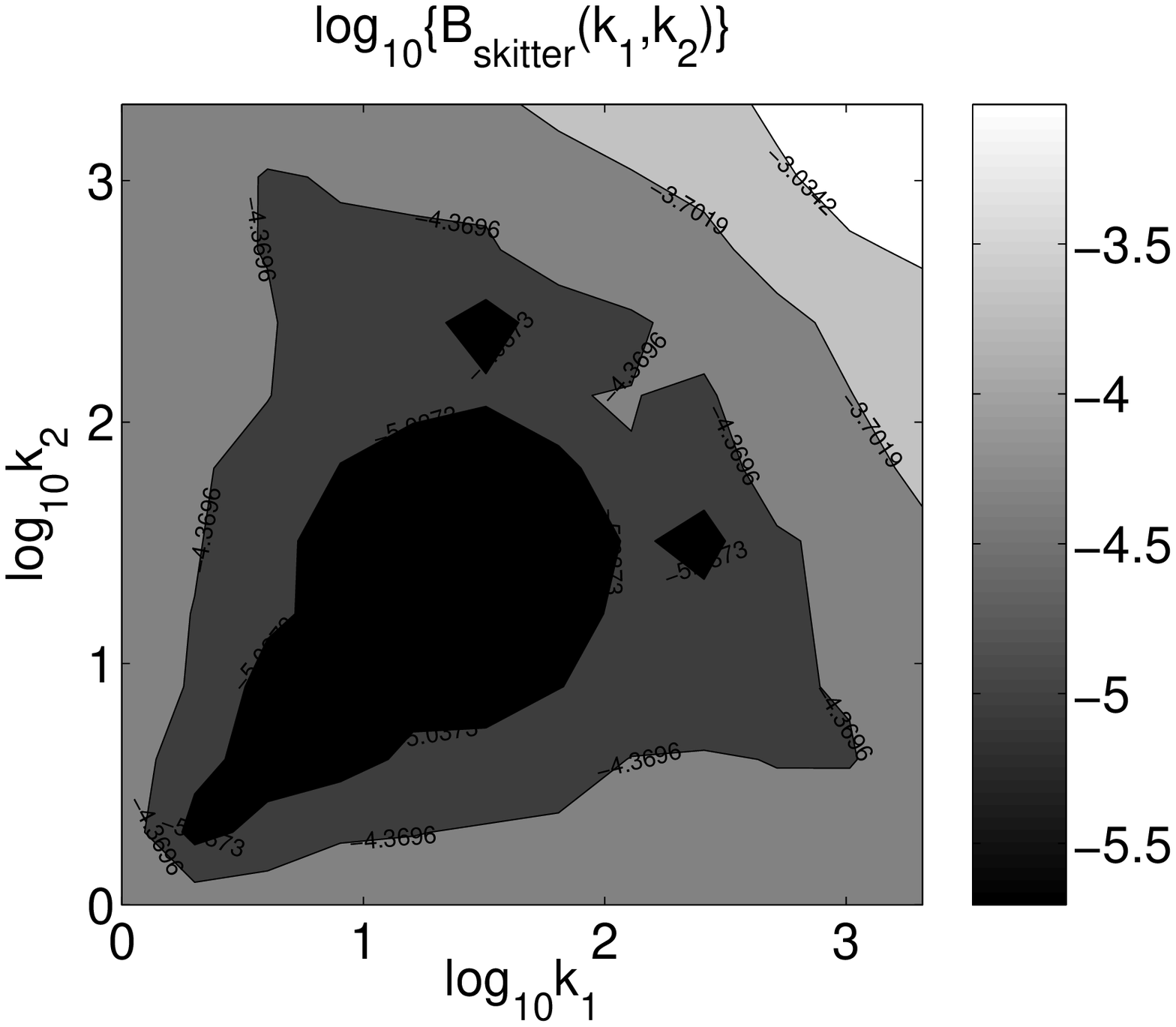}
        \label{fig:bkk-skitter}}
        \hfill
        \subfigure[BGP tables]
        {\includegraphics[width=2in]{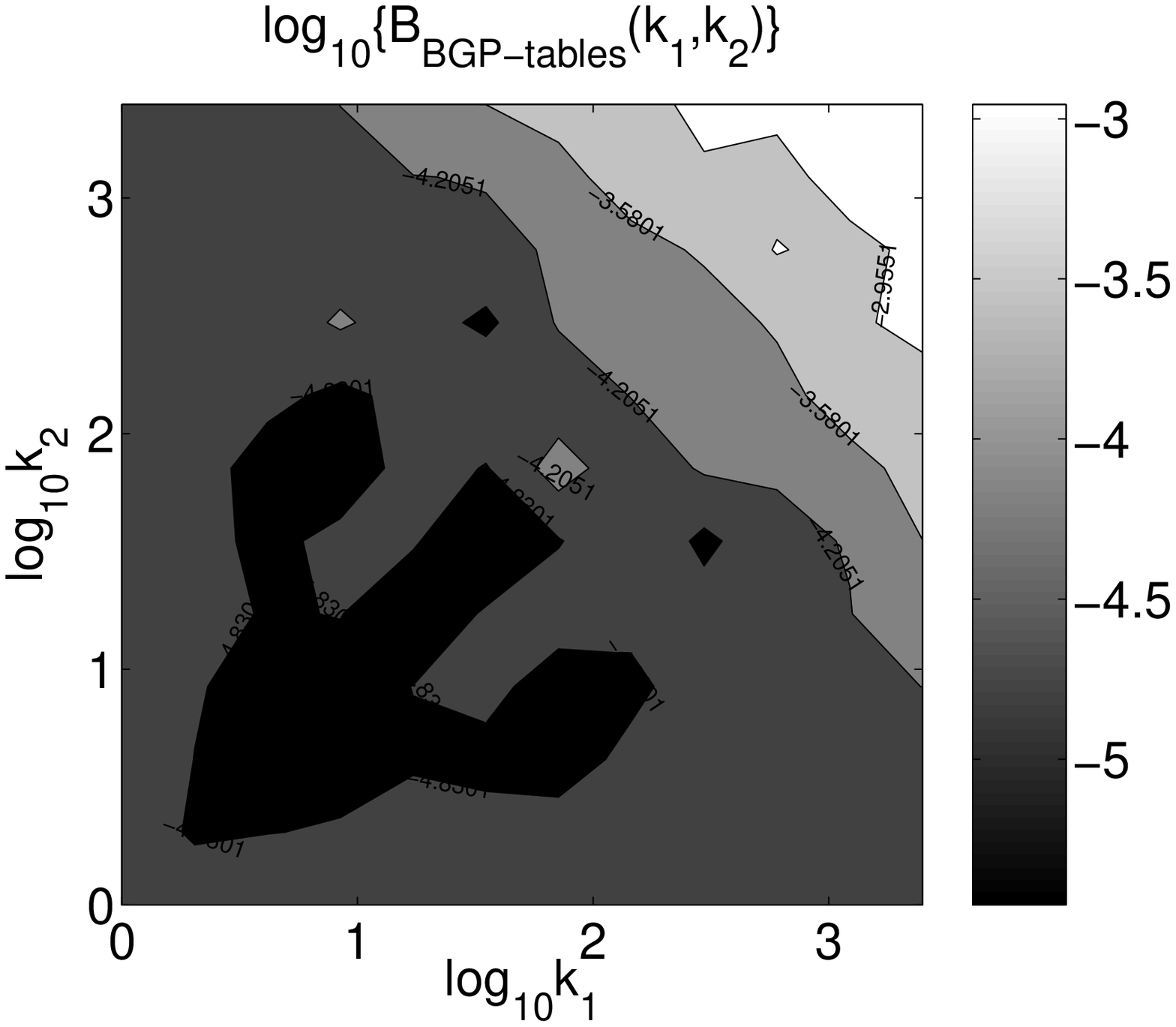}
        \label{fig:bkk-bgp}}
        \hfill
        \subfigure[WHOIS]
        {\includegraphics[width=2in]{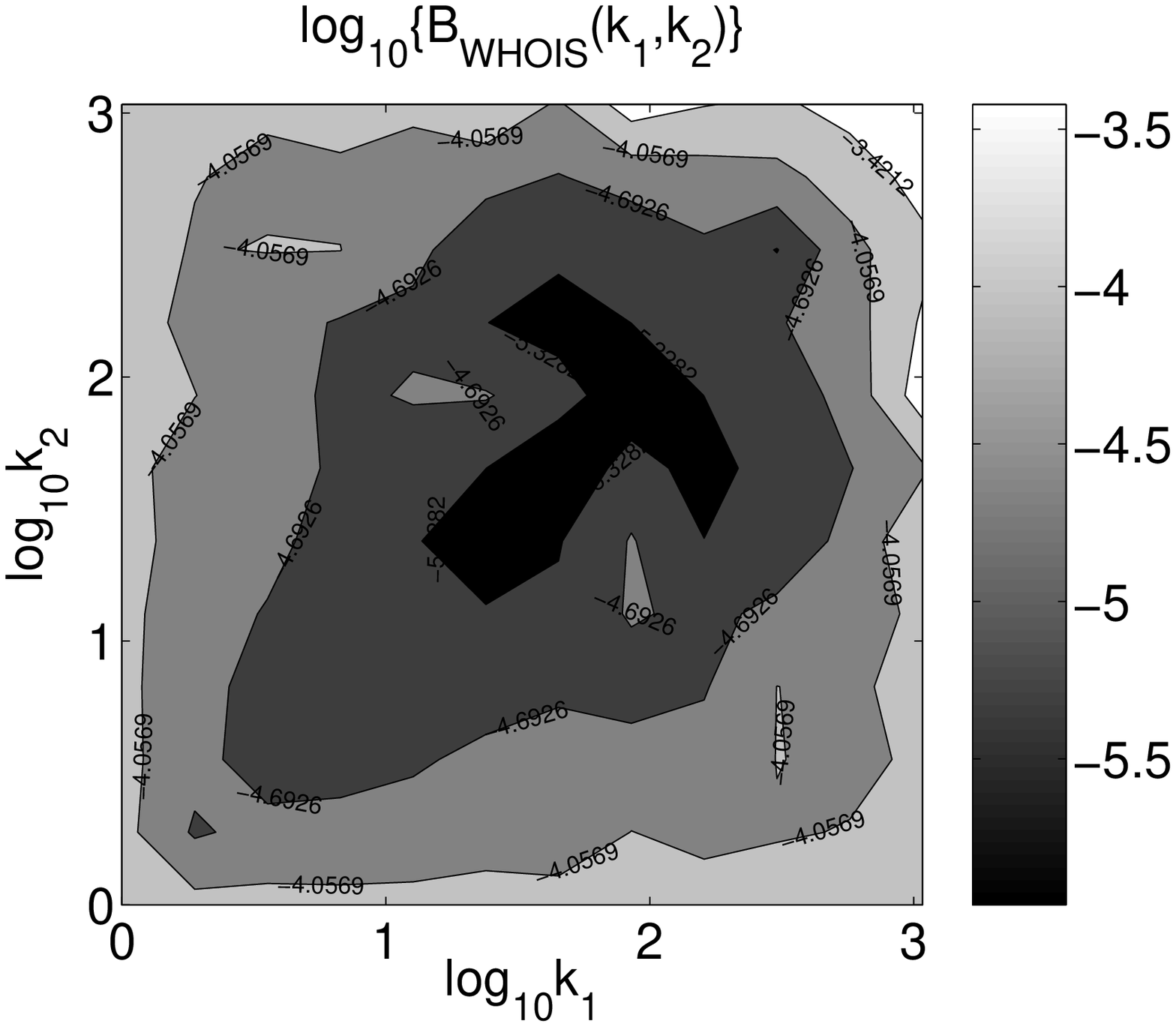}
        \label{fig:bkk-whois}}
    }
    \caption{\footnotesize {\bf Logarithm of normalized link
    betweenness~$\mathbf{B(k_1,k_2)/n/(n-1)}$ on a log-log scale.}}
    \label{fig:bkk}
\end{figure*}

\textbf{\textit{Importance.}}
Betweenness is important for traffic engineering applications that
try to estimate potential traffic load on nodes/links and potential congestion
points in a given topology. Betweenness is also critical for evaluating the
accuracy of topology sampling by tree-like probes (e.g.~{\em skitter} and BGP).
As shown in~\cite{DaAlHaBaVaVe05}, the
broader the betweenness distribution, the higher the statistical accuracy of the
sampled graph. The exploration process statistically focuses on
nodes/links with high betweenness thus providing an accurate sampling of the
distribution tail and capturing relevant statistical information. Finally we
note that {\em link value}, used~\cite{TaGoJaShWi02} to analyze the topology
hierarchy, and {\em router utilization}, used~\cite{LiAlWiDo04} to measure
network performance, are both directly related to betweenness.

\textbf{\textit{Discussion.}}
The simplest approach to calculating node betweenness requires long run
times, but we used an efficient algorithm from~\cite{brandes01}. We had to modify it
to also compute link betweenness.

For skitter and BGP graphs,
node betweenness is a growing power-law function of node degree
(Figure~\ref{fig:betweenness}) with exponents given in Table~\ref{table:summary}.
An excess of medium degree nodes in the WHOIS graph (Figure~\ref{fig:pk})
leads to greater path diversity and, hence, to lower betweenness values for
these nodes.

We also calculate average link betweenness as a function of degrees
of nodes adjacent to a link~\mbox{$B(k_1,k_2)$} (Figure~\ref{fig:bkk}).
The contour plots provide information on
the betweenness values of the links that connect similar or dissimilar degree
nodes. One would expect links connecting high-degree nodes to exhibit
highest link betweenness and thereby be used as a measure of link centrality.
Contrary to popular belief, the contour plots show that link betweenness
does not measure link centrality.
First, betweenness of links adjacent to low-degree nodes
(the left and bottom sides of the plots) is not the minimum. In fact,
non-normalized betweenness of links adjacent to 1-degree nodes is constant
and equal to~\mbox{$n-1$} (the number of destinations in the rest of the
network). Similar values of betweenness characterize links elsewhere
in the graph, including radial links between high and low-to-medium degree
nodes and tangential links in the zone of medium-to-high degrees (diagonal zone
from bottom-right to upper-left).
While the maximum-betweenness links are between high-degree nodes as expected
(the upper right corner of the plots), the minimum-betweenness links are tangential in the medium-to-low degree zone
(diagonal areas of low values from bottom-left to upper-right). We can explain
the latter observation by the following argument. Let~$i$ and~$j$ be two nodes
connected by a minimum-betweenness link~$l$. The only shortest paths going
through~$l$ are those between nodes that are {\em below}~$i$ and~$j$, where
``below'' means further from the core and closer to the fringe. When the
degrees of both~$i$ and~$j$ are small, the numbers of nodes below them (with
lower degree) are small, too. Consequently, the number of shortest paths,
proportional to the product of the number of nodes below~$i$ and~$j$, attains
its minimum at~$l$. We conclude that link betweenness is not a measure of
centrality but a measure of a certain combination of link centrality and radiality.

\subsection{Spectrum}\label{sec:characteristics:spectrum}

\textbf{\textit{Definition.}}
Let~$\mathcal{A}$ be the
adjacency matrix of a graph. This~\mbox{$n \times n$}
matrix is constructed by setting the value of its
element~\mbox{$a_{ij}=a_{ji}=1$} if there is a link between nodes~$i$ and~$j$.
All other elements have value~$0$. Scalar~$\lambda$ and vector~$v$ are
the eigenvalue and eigenvector respectively of~$\mathcal{A}$
if~\mbox{$\mathcal{A} v = \lambda v$}.
The {\bf spectrum} of a graph is the set of eigenvalues of its
adjacency matrix.

Another closely related and frequently used
definition of the graph spectrum is the spectrum of the eigenvalues of
its Laplacian,
$\mathcal{L} = \mathcal{T}^{-1/2}(\mathcal{T}-\mathcal{A})\mathcal{T}^{-1/2}$,
where~$\mathcal{T}$ is the diagonal matrix with~$t_{ii}$ equal to the degree
of node~$i$. This definition
is a normalized version of the original definition, in the sense that
for any graph, all the eigenvalues of its Laplacian are located
between~$0$ and~$2$. We use the original definition in this paper.

\begin{figure}[tbh]
      \centerline{
          \includegraphics[width=2.5in]{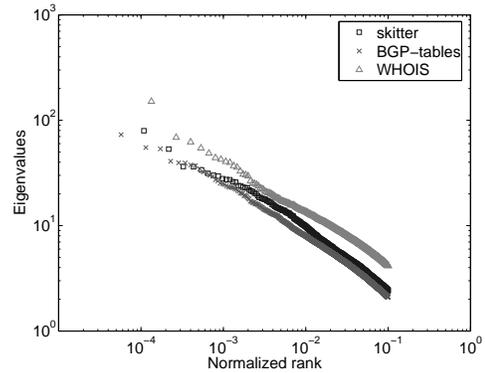}
      }
      \caption{\footnotesize {\bf Spectrum.} Absolute values of top 10\% of
      eigenvalues ordered by their normalized rank: normalized rank is node rank divided
      by the total number of nodes in the graph.}
      \label{fig:spectrum}
\end{figure}
\textbf{\textit{Importance.}}
Spectrum is one of the most important
{\em global\/} characteristics of the topology.
Spectrum yields tight bounds for a wide range of critical graph
characteristics~\cite{chung97}, such as distance-related parameters, expansion
properties, and values related to separator problems estimating graph resilience
under node/link removal. The largest eigenvalues are particularly important.
Most networks with high values for these largest eigenvalues have small
diameter, expand faster, and are more robust.

Two specific examples of spectrum-related metrics that made significant
contributions to networking topology research further emphasize the importance
of spectrum.
First, Tangmunarunkit {\it et al.}~\cite{TaGoJaShWi02} defined network {\em resilience},
one of the three metrics critical for their topology comparison analysis, as a measure
of network robustness under link removal, which equals the minimum balanced cut size of
a graph. By this definition, resilience is related to spectrum since the graph's largest
eigenvalues provide bounds on network robustness with
respect to both link {\em and\/} node removals~\cite{chung97}.

Second, Li {\it et al.}~\cite{LiAlWiDo04} define network {\em performance}, one of the two
metrics critical for their HOT argument, as the maximum traffic throughput of the network.
By this definition, performance is related to spectrum since it is essentially the
network conductance~\cite{GkaMiSa03} which can be tightly estimated by the gap between the
first and second largest eigenvalues~\cite{chung97}.

Beyond its significance for network robustness and performance,
the graph's largest eigenvalues are important
for traffic engineering purposes since graphs with larger eigenvalues
have, in general, more node- and link-disjoint paths to choose from.
The spectral analysis of graphs is a powerful tool for detailed investigation
of network structure, such as discovering clusters of highly interconnected
nodes~\cite{VuHuEr01}, and possibly revealing the hierarchy of ASes in the
Internet~\cite{GkaMiZe03}.

\textbf{\textit{Discussion.}}
Our $\bar{k}$-order (BGP, skitter, WHOIS)
plays a key role once again: the densest graph, WHOIS, is at the top in
Figure~\ref{fig:spectrum} and its first eigenvalue is largest in Table~\ref{table:summary}.
The eigenvalue distributions of all the three graphs follow power laws.

%% file: conclusion.tex
We presented a detailed comparison of widely available sources
of Internet topology data---skitter, BGP, and WHOIS---in terms
of a number of popular metrics studied in the literature.
Of the set of metrics we considered, the {\em joint degree
distribution\/}~(JDD)~\mbox{$P(k_1,k_2)$} appears to play
a central role in determining a wide range of other topological properties.
Indeed, using only the average degree~$\bar{k}$ and
the assortativity coefficient~$r$,
the two coarse summary statistics of the JDD, we could explain the
relative order of all other metrics for all our data sources.
At the same time, we saw that the values of~$\bar{k}$ and~$r$
are closely connected with the data source properties and
collection methodologies.
While additional work is required to assess the
definitiveness of the JDD in describing topologies, we have demonstrated
that it is a powerful metric for
capturing a variety of important graph properties. Isolating such an
encompassing
metric or a small set of metrics is a prerequisite to developing
accurate topology generators since it would reduce the
number of parameters one has to reproduce.
Building a JDD-based topology generator and
investigating the roles of degree correlations of higher orders
are subjects of our current research.

A number of methods have been proposed~\cite{SuAgReKa02,DiKrHuClRi05}
to annotate links in AS-level graphs thus incorporating AS relationship
information. Although we did not consider AS relationships in this study,
we note that the results of our analysis, in general, and JDD-related
statistics, in particular, are immediately applicable to directed---or, more
generally, {\em annotated}---graphs as well.

It remains an open question which data source most closely matches actual
Internet AS topology, given that each graph approximates a different view
of the Internet looking at the data (skitter), control (BGP), and
management (WHOIS) planes. In particular, we want to know what data source
contains reliable information about what type of links and how over-
or under-reporting of such links affects the metric values in the resulting
graphs. This knowledge would allow us to combine
information that we trust from all three data sources so that
we can obtain the most representative and complete Internet
topology view. For now, we see that topologies derived from the three data
sources are
quantitatively but not qualitatively different:
all three degree distributions are scale-free, but not all of them are
power laws. We conclude that comparative analysis of these three views allows
us to test the limits of metrics' sensitivity
to measurement incompleteness and inaccuracies.

We believe that our work will arm researchers with deeper insights
into specifics of each topology view. We hope that this study
encourages the validation of existing topology models against real
data and motivates the development of better ones.

\begin{table*}[tbh]
    \centering
    \caption{\bf Summary statistics.
     }
    \input{tables/connectivity.tex}
    \label{table:summary}
\end{table*}

%% file: tables/connectivity.tex
\begin{tabular}{|c|c|c|c|c|}
\cline{3-5}
 \multicolumn{2}{c|}{ }  & skitter  & BGP tables  & WHOIS \\ \hline
 Average degree & Number of nodes {\footnotesize $(n)$ } & 9,204 & 17,446 & 7,485 \\ \cline{2-5}
  & Number of edges {\footnotesize $(m)$ } & 28,959 & 40,805 & 56,949 \\ \cline{2-5}
  & Avg node degree {\footnotesize  $(\bar{k})$ } & 6.29 & 4.68 & 15.22 \\ \cline{2-5}
\hline
 Degree distribution & Max node degree  {\footnotesize $(k_{max})$ } & 2,070 & 2,498 & 1,079 \\ \cline{2-5}
  & Power-law max degree { \footnotesize $(k_{max}^{PL}) $ } & 1,448 & 4,546 & - \\ \cline{2-5}
  & Exponent of {\footnotesize $P(k)\;(-\gamma)$ } & 2.25 & 2.16 & - \\ \cline{2-5}
\hline
 Joint degree distribution & Avg neighbor degree {\footnotesize $(\bar{k}_{nn}/(n-1))$ } & 0.05 & 0.03 & 0.02 \\ \cline{2-5}
  & Exponent of {\footnotesize $k_{nn}(k)\;(-\gamma_{nn})$ } & 1.49 & 1.45 & - \\ \cline{2-5}
  & Assortative coefficient {\footnotesize $(r)$ } & -0.24 & -0.19 & -0.04 \\ \cline{2-5}
\hline
 Clustering & Mean clustering {\footnotesize $(C_{mean})$ } & 0.46 & 0.29 & 0.49 \\ \cline{2-5}
  & Clustering coefficient {\footnotesize $(C_{coeff})$ } & 0.03 & 0.02 & 0.31 \\ \cline{2-5}
  & Exponent of {\footnotesize $C(k)\;(-\gamma_{C})$} & 0.33 & 0.34 & - \\ \cline{2-5}
\hline
Rich club  & Exponent of {\footnotesize $\phi(\rho/n)\;(-\gamma_{rc})$} & 1.48 & 1.45 & 1.69 \\ \cline{2-5}
\hline
 Distance & Avg distance {\footnotesize $(\bar{d})$} & 3.12 & 3.69 & 3.54 \\ \cline{2-5}
  & Std deviation of distance {\footnotesize $(\sigma)$} & 0.63 & 0.87 & 0.80 \\ \cline{2-5}
  & Exponent of {\footnotesize $d(k)\;(-\gamma_d)$} & 0.07 & 0.07 & 0.09 \\ \cline{2-5}
\hline
 Betweenness & Avg node betweenness {\footnotesize $(\bar{B}_{node}/(n(n-1)))$} & $11 \cdot 10^{-5}$  & $7.7 \cdot 10^{-5} $ & $17 \cdot 10^{-5}$ \\ \cline{2-5}
  & Exponent of {\footnotesize $B(k)\;(\gamma_B)$ } & 1.35 & 1.17 & - \\ \cline{2-5}
  & Avg edge betweenness  {\footnotesize $(\bar{B}_{edge}/(n(n-1)))$ } & $5.37 \cdot 10^{-5} $ & $4.51 \cdot 10^{-5} $ & $3.10 \cdot 10^{-5} $ \\ \cline{2-5}
\hline
 Spectrum & Largest eigenvalue & 79.53 & 73.06 & 150.86 \\ \cline{2-5}
  & Second largest eigenvalue & -53.32 & -55.13 & 68.63 \\ \cline{2-5}
  & Third largest eigenvalue & 36.40 & 53.54 & 62.03 \\ \cline{2-5}
\hline
\end{tabular}

%% file: ack.tex
We thank Ulrik Brandes for sharing his betweenness code with us,
Andre Broido for answering our questions, and anonymous reviewers
for their comments that helped to improve this paper.

Support for this work was provided by NSF CNS-0434996,
NCS ANI-0221172, Center for Networked Systems (UCSD), and Cisco University
Research program.